\documentstyle[12pt,aaspp4]{article}
\everymath{\rm}
\everydisplay{\sf}
\received{}
\revised{}
\accepted{}
\cpright{}{}\journalid{}{}
\articleid{}{}
\paperid{}
\slugcomment{To appear in {\it The Astrophysical Journal}}
\begin{document}
\title{On The Cosmic Origins Of Carbon And Nitrogen}

\author{R.B.C. Henry\footnote{Department of Physics \& Astronomy, University of Oklahoma, Norman, OK 73019, USA; henry@mail.nhn.ou.edu}, M.G. Edmunds\footnote{Department of Physics \& Astronomy, Cardiff University, P.O. Box 913, Cardiff, CF2 3YB Wales; mge@astro.cf.ac.uk}, and J. K{\"o}ppen\footnote{International Space University, Blvd. Gonthier d'Andernach, F-67000 Illkirch, France}$^,$\footnote{Institut f{\"u}r theor. Physik u. Astrophysik der Universit{\"a}t zu Kiel, 24098 Kiel, Germany}$^,$\footnote{Observatoire Astronomique, UMR 7550, 11 Rue de l'Universite\'e, F-67000 Strasbourg, France; koppen@astro.u-strasbg.fr}}

\begin{abstract}

We analyze the behavior of N/O and C/O abundance ratios as a function of metallicity as gauged by O/H in large, extant Galactic and extragalactic H~II region abundance samples. We compile and compare published yields of C, N, and O for intermediate mass and massive stars and choose appropriate yield sets based upon analytical chemical evolution models fitted to the abundance data. We then use these yields to compute numerical chemical evolution models which satisfactorily reproduce the observed abundance trends and thereby identify the most likely production sites for carbon and nitrogen. Our results suggest that 
carbon and nitrogen originate from separate production sites and are decoupled from one another.
Massive stars (M$>$8~M$_{\sun}$) dominate the production of carbon, while intermediate-mass stars between 4 and 8~M$_{\sun}$, with a characteristic lag time of roughly 250~Myr following their formation, dominate nitrogen production.  Carbon production is positively sensitive to metallicity through mass loss processes in massive stars and has a pseudo-secondary character. Nitrogen production in intermediate mass stars is primary at low metallicity, but when 12+log(O/H)$>$8.3, secondary nitrogen becomes prominent, and nitrogen increases at a faster rate than oxygen -- indeed the dependence is steeper than would be formally expected for a secondary element.
The observed flat behavior of N/O versus O/H in metal-poor galaxies is explained by invoking low star formation rates which flatten the age-metallicity relation and allow N/O to rise to observed levels at low metallicities. The observed scatter and distribution of data points for N/O challenge the popular idea that observed intermittent polluting by oxygen is occurring from massive stars following star bursts. Rather, we find most points cluster at relatively low N/O values, indicating that scatter is caused by intermittent increases in nitrogen due to local contamination by Wolf-Rayet stars or luminous blue variables. In addition, the effect of inflow of gas into galactic systems on secondary production of nitrogen from carbon may introduce some scatter into N/O ratios at high metallicities. 

\end{abstract}


\clearpage

\section{Introduction}

``It is quite a three pipe problem'' (S. Holmes, quoted in Doyle 1891).

Carbon and nitrogen are among the most abundant of the chemical elements, and of obvious importance for life. Carbon is also a major constituent of interstellar dust. The main nuclear processes which generate these two elements are reasonably well understood -- the carbon must come predominantly from the triple-alpha reaction of helium, and nitrogen by the conversion of carbon and oxygen that occurs during the CNO cycles of hydrogen burning. A lingering problem, though, has been the lack of knowledge of which {\it sites} are most important for their generation -- in particular do they come mainly from short-lived massive stars, or from longer-lived progenitors of asymptotic giant branch stars? Coupled with this is uncertainty over the form and magnitude of the dependence of their production on the metallicity of the stars in which the reactions take place. Metallicity can essentially be tracked by its principal component -- oxygen -- whose dominant source is the Type~II supernova explosion of massive stars. There is observed variation in the ratios of carbon and
 nitrogen to oxygen (i.e. C/O, N/O) in both stars (e.g. Gustafsson et al. 1999) and the gas in galactic systems (e.g. Garnett et al. 1999; Henry \& Worthey 1999), and it is these variations that we use as clues to pin down the synthesis sites.

The threshold temperature of He burning and production of $^{12}$C via the triple alpha process is $\sim$10$^8$K, a temperature accessible in both massive (M$>$8M$_{\sun}$) and intermediate mass (1$<$M$<$8M$_{\sun}$) stars. Thus, these broad stellar groups represent two possible sites for carbon production. Likewise, nitrogen production via the CNO cycles may occur in either of these sites. However, discovering the origin of nitrogen is further complicated by the fact that the seed carbon needed for its production may either have been present when the star was born or is synthesized within the star during its lifetime. We now explore this idea in more detail.

Nitrogen is mainly produced in the six steps of the CN branch of the
CNO cycles within H burning stellar zones, where $^{12}$C serves as
the reaction catalyst (see a textbook like Clayton 1983 or Cowley 1995
for nucleosynthesis review).  Three reactions occur to transform
$^{12}$C to $^{14}$N: $^{12}$C(p,$\gamma$)$^{13}$N($\beta$$^{+}$,$
\nu$)$^{13}$C(p,$\gamma$)$^{14}$N, while the next step, 
$^{14}$N(p,$\gamma$)O$^{15}$, depletes nitrogen and has a
relatively low cross-section. The final two reactions in the
cycle transform $^{15}$O to $^{12}$C. Since the fourth reaction
runs much slower than the others, 
the cycle achieves equilibrium only when $^{14}$N accumulates to high
levels, and so one effect of the CN
cycle is to convert $^{12}$C to $^{14}$N. The real issue in
nitrogen evolution is to
discover the source of the carbon which is converted into nitrogen, and of
any oxygen which can contribute through the (slow) side chain $^{16}$O(p,$\gamma$)$^{17}$F($\beta$$^{+}$,$\nu$)$^{17}$O(p,$\alpha$)$^{14}$N.

The conventional meaning of ``{\it primary}'' applied to nitrogen is that its
production is independent of the initial composition of the star in which
it is synthesized. An example is where stars produce their own carbon (and 
some oxygen) during helium burning, and the carbon (and perhaps oxygen) is
subsequently processed into $^{14}$N via the CN(O) cycle. Stars beyond the
first generation in a galactic system already contain some carbon and oxygen,
inherited from the interstellar medium out of which they formed. The amount
of nitrogen formed from CNO cycling of this material will be proportional to
its C abundance (and also its O abundance, if the CNO cycling proceeds long
enough to deplete the oxygen) and is known as ``{\it secondary}'' nitrogen. 
In general, then, primary nitrogen production is independent of metallicity, while secondary production is a linear function of it.

However, these conventional definitions become rather blurred if the
evolution of the synthesizing stars, and/or the release of nucleosynthesis
products to the interstellar medium, depends on the initial composition
of the star - as well it might if stellar wind generation is metallicity-dependent. Thus effective production of carbon could depend on the initial metallicity of the star, although no actual ``seed nucleus'' is involved, and the production of nitrogen might differ from a simple primary or secondary process. Also
important is the mass of star in which production takes place, since if 
{\it most} production takes place in stars of {\it low} mass, a significant
delay will occur between formation of the source stars and release of the 
products into the interstellar medium.

Detailed discussion and review of computed stellar yields is left until
{\S\S}3 and 5 below, but we now refer to current {\it interpretations} of the observed carbon and nitrogen abundances in stars and galaxies. The literature on the
origin of carbon has recently been reviewed by Gustafsson et al. (1999),
and Garnett et al. (1999). The former conclude that their own stellar 
results are ``consistent with carbon enrichment by superwinds of metal-rich
massive stars, but inconsistent with a main origin of carbon in low mass 
stars'', which is pretty much echoed by the latter authors who state that the 
behavior of C/O ratios as a function of O/H is ``best explained [by] ...
effects of stellar mass loss on massive star yields''. They note that
theoretical chemical evolution models (Carigi 1994) in which carbon comes
from intermediate stars apparently predict too shallow a C/O relation to fit
their observed galaxy abundance gradients. Other recent discussions of C/O 
gradients across our own or other galaxies, or of C/O ratios in low 
metallicity galaxies, have been given by G\"{o}tz and K\"{o}ppen (1992), 
Prantzos, Vangioni-Flam \& Chauveau (1994), Kunth et al. (1995),
Garnett et al. (1995), Carigi et al. (1995), Moll\'{a}, Ferrini,
and D{\'i}az (1997) and Chiappini, Matteucci \& Gratton (1997).

The literature on the origin of nitrogen was briefly reviewed by Vila-Costas
\& Edmunds (1993). A major problem with N/O ratios has been to try to explain
the {\it spread} in N/O at a given O/H (see Figure 1B of Section 2), although
the reality of a spread at low O/H (12+log(O/H) $\leq$ 7.6) has been 
questioned by Thuan et al. (1995) and Izotov \& Thuan (1999). Two major
mechanisms have been proposed for generating a spread - mechanisms that could
also apply (with different timescales etc) to C/O ratios. One mechanism invokes a significant time delay between formation of the star which will
produce the nitrogen and the delivery of the nitrogen to the interstellar
medium. Thus oxygen is expected to be produced predominantly in the SNII
explosion of short-lived massive stars, the N/O ratio in the ISM will at first
decrease, and then rise again as the nitrogen is released. A delay could affect
both primary and secondary nitrogen, and Edmunds \& Pagel (1978) suggested 
that N/O ratios might perhaps be an indicator of the age of a galactic system,
in the sense that it indicated the time since the bulk of star formation has
taken place. This idea continues to find some support (e.g. Kobulnicky \&
Skillman 1996; van Zee, Haynes \& Salzer 1997). A suspicion of low 
N/S or N/Si ratios (the S and Si being expected to follow O) in damped Lyman
alpha absorption systems (Lu et al. 1996; Pettini, Lipman \& Hunstead 1995; Pilyugin 1999)
has been invoked as evidence of the youth of the systems on the basis of
delayed N release. However, if the delay mechanism is to be effective in
altering N/O ratios, the delay must be reasonably long - otherwise the probability of catching a system at low N/O, perhaps after a burst of star formation, will be too small. A time scale of {\it several} 10$^{8}$ or of order 10$^{9}$ 
years would seem to be necessary. We shall argue later that the dominant source of nitrogen may always be in 
intermediate mass stars of too high a mass to allow a strong, observable, systematic effect.

The second mechanism for causing N/O variation (or C/O variation, but to avoid
repetition we only discuss nitrogen here) is variation in the flow of gas into
or out of the galaxy. If the nitrogen is {\it primary}, then neither inflow
of unenriched gas nor outflow of interstellar medium will affect the N/O ratio
{\it except} in the case where the outflow is {\it different} for nitrogen and
oxygen (e.g. Marconi, Matteucci \& Tosi 1994; Pilyugin 1993). If the nitrogen
is {\it secondary} (or of any composition behavior other than primary) the
N/O ratio is still unaffected by any {\it non-differential} outflow, but can
be affected by unenriched inflow (Serano \& Peimbert 1983; Edmunds 1990).
K{\"o}ppen \& Edmunds (1999) were able to place the useful constraint that
variation in N/O caused by inflow and secondary nitrogen can be at most a 
factor of two if the inflow is time-decreasing (as most chemical evolution models tend to assume).

We will find - with no surprise - that nitrogen can be interpreted as having both primary and secondary components. But we will suggest that the nitrogen is
(or acts as if it is) secondary on {\it carbon}, rather than on oxygen. This 
will allow a rather steep dependence of N/O on O/H at high O/H, possibly aided
by inflow effects. That the CNO cycle might not go to completion, but effectively stop at CN equilibrium, was noted in LMC planetary nebulae by Dopita et al. (1996) and (at low metallicity) from observations of old stars by Langer \& Kraft (1984). We shall not discuss isotope ratios such as $^{15}$N/$^{14}$N
here, although they are a subject of some recent interest (Chin et al. 1999; Wielen \& Wilson 1997).
   
After reviewing the observational data in Section 2 and stellar yields in Section 3, we give an elementary analytic model which can account well for the general trends of the observational data. The yield parameters of this analytic model can then be compared with published stellar evolution and nucleosynthesis predicted yields with the result that it is possible to choose which are the most realistic yield calculations -- and to identify the stellar sources of carbon and nitrogen. The general form of the analytic models is confirmed by calculations generated by detailed chemical evolution codes in Section 5. After some discussion in Section 6, our identification of the sources and metallicity-dependence of carbon and nitrogen production are summarized in {\S}7.

\section{Data}

We begin by considering the observed trends of carbon and nitrogen
abundances with metallicity as gauged by oxygen abundance.  Abundance
data used for our study were taken directly from the literature, and the
references are summarized in Table~1, where we indicate the type of object
observed along with the spectral range, the first author of the study, and the
number of objects.  The first five studies pertain to objects within the Milky Way only,
while the rest refer to objects in external galaxies.  The data for carbon and
nitrogen are presented in Figs.~1A and 1B, respectively.  Below we discuss the
data sources and the observed trends.

Carbon is an element whose abundance has lately become more measurable in
extra-galactic H~II regions, thanks to the Hubble Space Telescope and its
UV capabilities, since the strong carbon lines of C~III] and C~IV appear in that
spectral region.  Fig.~1A is a plot of log(C/O) versus 12+log(O/H) for numerous
Galactic and extragalactic objects.  Results for extragalactic H~II regions are
taken from Garnett et al.  (1995; 1997; 1999), Izotov \& Thuan (1999), and Kobulnicky \& Skillman (1998),
and are shown with symbols `G', `I', and `K', respectively.  The filled circles
correspond to stellar data from Gustafsson et al.  (1999) for a sample of F and
G stars, the filled boxes correspond to B~star data from Gummersbach et al. (1998), and the filled diamonds are halo star data points from Tomkin et al. (1992).   The Galactic H~II regions M8 and the Orion Nebula have been measured
by Peimbert et al.  (1993) and Esteban et al.  (1998), respectively, where the
point for Orion is indicated with an `O' and M8 with an `M'.  The sun's position
(Grevesse et al. 1996) is shown with an `S'. Garnett et al. (1999) calculated two C/O ratios each for six objects, using two different reddening laws. The common values of 12+log(O/H) for these pairs are: 8.06, 8.16, 8.39, 8.50, 8.51, and 8.58. We note that the two points at the lowest 12+log(O/H) values in the Garnett et al. and Izotov \& Thuan samples are for I~Zw~18.

Data in Fig.~1A suggest a direct correlation between C/O and O/H which
has been noted before (c.f.  Garnett et al. 1999), although the result
is weakened somewhat by the two points for I~Zw~18 around
12+log(O/H)=7.25 along with the data for the halo stars.  Assuming that the trend is robust, it clearly
implies that carbon production is favored at higher metallicities.  One
promising explanation (Prantzos, Vangioni-Flam, \& Chauveau 1994;
Gustafsson et al. 1999) is that mass loss in massive stars is enhanced
by the presence of metals in their atmospheres which increase the UV
cross-section to stellar radiation. Stellar yield calculations by
Maeder (1992) appear to support this claim. The contributions to carbon
production by different stellar mass ranges is discussed by both
Prantzos et al. and  Gustafsson et al., who conclude that the bulk of carbon is 
produced by massive stars.  However, it is also
clear that stars less massive than about 5M$_{\odot}$ produce
and expel carbon as well (van den Hoek \& Groenewegen 1997; Marigo et
al. 1996, 1998; Henry et al. 2000), and the relative significance of 
massive and intermediate mass stars has not been established.

Fig.~1B shows log(N/O) versus 12+log(O/H).  The numerous data sources are
identified in the figure caption.  The most striking feature in Fig.~1B is the
apparent threshold running from the lower left to upper right beginning around
12+log(O/H)=8.25 and breached by only a few objects. For the remainder of the paper, we shall refer to this threshold as the {\it NO envelope}.
Behind the NO envelope the
density of objects drops off toward lower values of 12+log(O/H) and higher
values of log(N/O).  A second feature is the apparent bimodal behavior of N/O. At values of
12+log(O/H)$<$8, N/O appears constant, a trend which seems to be
consistent with the upper limits provided by the damped Ly$\alpha$ objects of Lu
et al.  (1996; L) and with the observed abundances of blue compact galaxies found by Izotov \& Thuan (1999) at very low metallicity. Then, at 12+log(O/H)$>$8 N/O turns upward and rises steeply.  The bimodal behavior of the NO envelope was emphasized
by Kobulnicky \& Skillman (1996), and is consistent with the idea that nitrogen and oxygen rise in lockstep at low metallicities,
while nitrogen production becomes a function of metallicity at values of
12+log(O/H) greater than 8.3. In summary, the shape of the NO envelope and the scatter of points behind it represent two related problems in understanding nitrogen synthesis.  In our analysis, we address primarily the first of these problems, while mentioning the second one briefly and deferring a detailed discussion of it to a future paper.

We employ analytical ({\S}4) and numerical ({\S}5) models in order to understand
the observed trends in carbon and nitrogen buildup with that of oxygen.  An
important component of the numerical models, however, are the stellar yields which enter
into the calculations, and these are discussed in detail next in {\S}3.

\section{Yields}

In this section we compare yield predictions by various authors for those
elements in whose evolution we are interested, namely carbon, nitrogen, oxygen. {\it Furthermore, we state clearly at the outset that throughout our discussion we are only considering the most abundant isotope of each of these elements, i.e. $^{12}$C, $^{14}$N, and $^{16}$O}.  Here, we
present and compare modern yields which are available in the literature and
which can be used to calculate both analytical and numerical models for the
purpose of understanding the observations of carbon and nitrogen.  There are two
important production sites which we shall consider:  (1)~intermediate mass stars
(IMS; 1$\le$M$\le$8M$_{\sun}$), potentially important for both carbon and nitrogen synthesis; and (2)~massive stars (M$>$8M$_{\sun}$) which
are important sites for oxygen production and {\it perhaps} for carbon and nitrogen as well.

Recent IMS yield calculations have been carried out by
van~den~Hoek and Groenewegen (1997; VG) and Marigo, Bressan, \& Chiosi
(1996; 1998; MBC)\footnote{Because of their use of modern opacity values and more
detailed mass loss processes, the studies by these two teams supercede the earlier work
of Renzini \& Voli (1981). However, we do not mean to imply that current use of these latter yields necessarily gives a spurious result, given that uncertainties even for contemporary yield calculations are relatively large.}. These models include the conversion of
carbon to nitrogen at the base of the convective envelope during third
dredge-up, the process known as ``hot bottom burning''.  The quality of the yield
predictions by VG and Marigo et al. has recently been assessed empirically by
Henry, Kwitter, \& Bates (2000), who compared predictions of {\it in situ} planetary
nebula abundances from these yield studies with their observed abundances in
a sample of planetaries and found modest but encouraging agreement
between theory and observation.

Yields for massive stars which we shall consider are those by Maeder (1992),
Woosley \& Weaver (1995) and Nomoto et al.  (1997).  The last two teams include
element production from both quiescent as well as explosive burning stages. The Maeder calculations are unique in that they include
the effects of metallicity on mass loss. We do not consider any yields of Type~Ia supernovae in our study.

We compare theoretical yields by first defining the {\it integrated yield} of an
element $x$ as:  
\begin{equation} 
P_x \equiv \int_{m_{down}}^{m_{up}}mp_x(m)\phi(m) dm, 
\end{equation} 
where $p_x$(m) is the stellar yield,
$\phi$(m) is the initial mass function, and $m_{up}$, $m_{down}$ are, respectively, upper and lower limits to the mass range of all stars formed. Assuming a
Salpeter initial mass function (see {\S}5.1) and a stellar mass range from 0.1 to 120~M$_{\sun}$, $P_x$ is then the mass fraction of all stars formed
which is eventually expelled as new element $x$.

Our calculated integrated yields are given in Table~2A, where column~1 indicates the source of values for $p_x$(m) in eq.~1 as defined in the footnote, columns 2 and 3 show the upper and lower mass limits of the range of stars considered in each study [note these limits are not the same as the integration limits in eq.~1, which refers to all stars], $z$ indicates the metallicity for which the yields are relevant, and the last three columns give the integrated yields for carbon, nitrogen, and oxygen, respectively, according to eq.~1. 

The VG and MBC studies refer to intermediate-mass stars where only carbon and nitrogen yields are of interest to us, since these stars do not synthesize oxygen. We see that results from both studies suggest that, with increasing metallicity, the integrated carbon yield decreases  while the nitrogen yield increases. According to a recent study of thermally pulsing AGB models by Buell (1997), stars of relatively low metallicity have smaller total radii (higher surface gravity) and less effective mass loss, increasing the lifetime of the AGB phase and allowing these stars
to experience more 3rd dredge-up when carbon is mixed out into the envelope. As metallicity rises, then, carbon production drops off while secondary nitrogen production becomes significant. In the balance, then, as stellar metallicity goes up, the models predict that carbon production declines while nitrogen production rises. 

In the case of massive stars, Maeder (1992) predicts a sizable increase in carbon production with metallicity as the result of a mass loss process which is metallicity-sensitive. Carbon yields from Woosley \& Weaver and Nomoto predict significantly less carbon than Maeder and show no apparent sensitivity to metallicity.
Woosley \& Weaver predict higher nitrogen production than does Nomoto for solar metallicity; in addition, the Woosley \& Weaver results indicate that nitrogen production increases with metallicity. Note that Maeder did not include the contribution from supernova ejection, only from winds, in his calculation of nitrogen yields; the relevant numbers in Table~2A from his paper are therefore lower limits. 
Oxygen yields are predicted by Maeder to correlate inversely with metallicity while Woosley \& Weaver predict them to correlate directly, albeit weakly, with metallicity. We note that this inverse sensitivity in Maeder's results will be seen below to influence significantly the C/O and N/O behavior in our models.
We also point out that yield calculations for massive stars which account for stellar rotation effects have been carried out by Langer \& Henkel (1995) and Langer et al. (1997). Their predicted yields appear to be similar to those of Maeder for $^{12}$C and $^{14}$N.


\section{Analytical Models}

Our aim in this section is to fit reasonable analytical models to the data presented in {\S}2 by 
altering input yield parameters. These latter values are then used to select the 
most appropriate sets of published yields to be employed as input for our follow-up 
numerical models.

An inspection of C/O versus O/H abundances shown in Fig.~1A suggests that carbon 
production increases with metallicity. We will model this analytically by the 
formal assumption of a primary and secondary component to the carbon yield, but 
we must emphasize that this does not necessarily imply that carbon {\it is} a 
``secondary'' product of a varying initial seed. The metallicity dependence of 
the yield may well come about because stellar evolution and the release of 
nucleosynthesis products from a star are affected by its mass-loss, and this 
mass-loss can be metallicity dependent. For the results presented here, we will 
assume that there is no time delay in the delivery of carbon to the interstellar 
medium compared to the delivery of oxygen. (Our subsequent identification of
the major site of carbon synthesis with massive stars confirms the validity
of this assumption.) To allow for the effects of 
gas flow (Edmunds 1990) we note that {\it outflow} should have no effect on 
primary/primary or secondary/primary element ratios, but {\it accretion (i.e. 
inflow)} of unenriched gas can have some effect. As shown in K{\"o}ppen \& 
Edmunds (1999), useful limits on the effects of {\it time-decreasing} inflows 
can be set by considering the elementary ``linear'' inflow model in which the 
rate of gas accretion is equal to a constant times the star formation rate. The 
analytic models show that inflow 
can steepen the N/O relation, and - as inflow is frequently evoked in 
Galactic chemical evolution models as a mechanism for solving the G-dwarf 
problem - we allow some modest inflow in our numerical models.

Following the notation of Edmunds (1990, as modified by K{\"o}ppen \& Edmunds 
1999) for a system with gas mass $g$, gas metallicity $z$ (by mass -- and simply 
proportional to the oxygen abundance), carbon abundance $z_c$ (by mass), and a 
linear accretion rate $\alpha a$ times the star formation rate, then considering 
the change in element abundances in the interstellar medium when a mass $ds$ is 
formed into stars (from op. cit. Eq.~11 with 6):
\begin{equation}
\frac{dz}{ds} = \frac{p-az}{g}
\end{equation}
\begin{equation}
\frac{dz_c}{ds} = \frac{p_{pc}+p_{sc}z-az_c}{g},
\end{equation}
where the general metallicity yield is $p$, and that of carbon is 
$p_{pc}+p_{sc}z$, where $p_{pc}$ is the primary component yield and $p_{sc}$ the 
secondary. It has also been (with justification) assumed that $z, z_c \ll 1$. Note that $\alpha$ is that fraction of interstellar material going into a generation of star formation that remains locked up in long-lived stars and remnants. 
The parameter characterizing accretion obeys $0 \le a$, and for the ``simple'' 
closed-box model $a \Rightarrow 0$ (limiting forms are given in Appendix A).

Dividing the two equations gives
\begin{equation}
{{dz_c}\over{dz}}+{{az_c}\over{p-az}} = {{p_{pc}+p_{sc}z}\over{p-az}},
\end{equation}
which may be solved to give
\begin{equation}
z_c={{p_{pc}z}\over{p}}+{{pp_{sc}}\over{a^2}} \left\{ \frac{az}{p} + \left(1- 
\frac{az}{p} \right) \ln \left(1-\frac{az}{p} \right) \right\};
\end{equation}
suitable values to (eyeball) fit the data of Fig.~1A are $p=0.01$ (fixed by
requiring solar oxygen abundance for a simple model of the solar 
neighbourhood), 
$p_{pc}=0.0012$, $p_{sc}=0.9$, and $a=0.1$, as shown with a bold curve in Fig.~2A. We also show with faint curves
the simple models for stronger accretion, i.e. $a=0.5$, $a=0.9$, to indicate
the spread in N/O values that can be generated by different accretion rates. (Accretion of less than 0.1 produces curves which are nearly coincident with that for 0.1, so we do not show them in Fig.~2A.)
Note that 
$z_c/z$ and $z_n/z$ represent {\it mass} ratios, and must be multiplied by the 
relevant atomic weight ratios for comparison with observed abundance {\it 
number} ratios. As explained in K{\"o}ppen \& Edmunds (1999), the effect of 
different accretion rates on the ``secondary'' component is no more than a 
factor of two in the $z_c/z$ ratio at a given metallicity $z$, if the accretion 
is decreasing with time. It is clear that the model gives a reasonable account 
of the data, with the effects of different accretion rates allowing some (small) 
spread in the element ratios at high abundance.

The next step is to extend the model to nitrogen abundances. The proposal here 
is that the yield of nitrogen has both a primary and a secondary component -- 
{\it but the latter is (or behaves as if it were) secondary on the carbon 
abundance, rather than on the overall metallicity}. In nucleosynthesis terms 
this might occur because the CN cycle comes into equilibrium much faster than 
the ON cycle. Complete conversion of the C into N could take place long before 
much O is converted into N. So provided the CNO cycling stops at this point, the 
major source of secondary nitrogen will have come from carbon. For the primary 
component, it is of little consequence whether the N comes from CNO cycling 
of C or O, since the C and O will have to have been produced in the star 
by the triple alpha 
reaction and $^{12}C(\alpha,\gamma)^{16}O$ and hence be insensitive to the 
star's initial abundance of either C or O. Some support for the cycling only converting the carbon may come from interpreting $^{17}$O/$^{16}$O ratios in the interstellar medium if overproduction of $^{17}$O is to be avoided, unless the standard nuclear reaction rate for $^{17}$O(p,$\alpha$)$^{14}$N is seriously in error (see Edmunds 2000).
But it could also be the case that 
the stellar evolution (with mass loss) mimics the effect of nitrogen being 
secondary on carbon because the overall effect is that the resulting N/O has a 
steeper metallicity than the proportionality to $z$ expected for a secondary 
process on oxygen. Taking for the moment the simpler view of secondary 
processing on the carbon, then we have a nitrogen yield $p_{pn}+p_{sn}z_c$, and 
\begin{equation}
\frac{dz_n}{ds} = \frac{p_{pn}+p_{sn}z_c-az_n}{g},
\end{equation}
which may be combined with Eqs.~2 and 5, and solved to give
\begin{eqnarray}
z_n& =&\left( \frac{p_{pn}}{p} + \frac{p_{pc}p_{sn}}{ap} + 
\frac{p_{sc}p_{sn}}{a^2} \right) z + \left( \frac{p_{pc}}{p} + \frac{p_{sc}}{a} 
\right) \left( \frac{pp_{sn}}{a^2} \right) 
\left( 1- \frac{az}{p} \right) \ln \left( 1- \frac{az}{p} \right)\\
&& - \frac{pp_{sc}p_{sn}}{2a^3} \left( 1- \frac{az}{p} \right) 
\left[ \ln \left( 1- \frac{az}{p} \right) \right]^2.\nonumber
\end{eqnarray}
Assuming the values of $p$, $p_{pc}$, and $p_{sc}$ given above,
suitable fitting values to the data of Fig.~1B are $p_{sn}=0.00022$ and 
$p_{sn}=0.285$, as shown in Fig.~2B for the same accretion rate values as in 
Fig.~2A. As can be seen, this ``double secondary'' model gives the 
characteristic knee, and steeper N/O dependence as O/H increases. Again, 
variation in inflow rate would give rise to some spread at moderate-to-high O/H 
values. The limiting form of Eq.~7 for the simple closed-box model is given in 
Appendix~A, in which the ``tertiary'' or ``double secondary'' component due to 
the processing of carbon is more apparent.

Comparing our analytic fit yield coefficients ($p$, $p_{pc}$, $p_{sc}$, $p_{pn}$, $p_{sn}$) with the results of the detailed
stellar yield calculations in {\S}3 ($P_C$, $P_N$, $P_O$), we find that for carbon our analytical 
primary and secondary values agree closely with 
Maeder's yields, and that the analytical values for nitrogen are similar to 
the intermediate-mass star predictions of VG and MBC. 

An elementary analytic treatment of the possible effect of a time delay in 
the nitrogen production is given in Appendix B. For illustration, the N/O 
versus O/H track for a closed system with a particularly rapid rate of star 
formation is plotted as a dashed line in Fig.~A1B. 

To summarize our analytic conclusions, we propose that the observational data on 
carbon and nitrogen abundance ratios relative to oxygen suggest that carbon has 
components that behave {\it as if} they were primary and secondary -- the 
secondary behavior most likely being due to dependence of stellar evolution and 
mass loss on metallicity. The nitrogen shows both primary and secondary 
components -- the 
latter behaving {\it as if it were secondary on carbon abundance}, i.e. 
demonstrating a steeper dependence on metallicity than would be expected if it 
were secondary on oxygen. Again, some of this behavior may be due to dependence 
of stellar evolution and mass loss on metallicity, rather than a purely nuclear 
seed effect. Different rates of infall of unenriched gas in galactic systems 
could explain some modest spread in secondary contribution to C/O ratios and 
(slightly greater effect) N/O ratios at high O/H. {\it Our analytical model has 
also suggested that Maeder's massive star yields for carbon and oxygen are the 
most appropriate, while the yields of VG or MBC are needed for nitrogen.} We 
explore this in much greater detail in the next section, where we present 
results of numerical models which incorporate these yields in them.

\section{Numerical Models}

We now wish to use the implications and results concerning yields of the last two sections to calculate numerical models which will assist us in closely identifying the important production site(s) of carbon and nitrogen in the Universe.
Specifically, we want to quantify the relative contributions of intermediate-mass and massive stars to the total mass buildup of these two elements as a function of time. To do this, we have constructed a numerical model of a generic one-zone galactic region and allowed it to evolve while we keep track of the contributions of intermediate-mass and massive stars to carbon and nitrogen synthesis at each timestep. The galaxy's mass builds from zero by infall (accretion) occurring
at a time-varying rate, while the rate of star formation is determined by the gas fraction. We do not assume instantaneous recycling, and thus stars eject matter after a time lag appropriate for their birth masses. The amounts of new helium, carbon, nitrogen, and oxygen synthesized and ejected by stars is assumed to be a function of progenitor mass and metallicity, using adopted yields inferred from analytical studies in {\S}4. We then trace the buildup in the interstellar gas of each element with time. Obviously, our final results will be directly linked to our choice of yields.

The abundance patterns of C/O and N/O shown in Figs.~1A,B are the result of plotting abundance ratios determined in H~II regions and stars from a broad sample of galaxies, including the Milky Way. Therefore, in computing our numerical models to fit the data, we make the following reasonable assumptions: (1)~stellar yields are strictly functions of progenitor mass and metallicity, and as such are not influenced by a star's local environment; the yields are universal and do not vary from galaxy to galaxy; (2)~likewise, the initial mass function is universal; and (3)~the star formation rate is a function of environment, since it is related to the local density and gas fraction. 
 We now describe the details of our numerical code. 

\subsection{Details Of Code}

Our numerical code for chemical evolution is a one-zone program written by one of us (R.B.C.H.)
which, for specified infall and star formation rates, follows
the buildup of carbon, nitrogen, and oxygen over time. The code utilizes the
formalism for chemical evolution described in Tinsley (1980),
and we now show and describe the relevant equations.

We imagine our generic galactic region as an open box, originally empty but accreting metal-free gas. As this
material accumulates and forms stars, the total mass of the region becomes partitioned into interstellar gas of mass $g$ and stars of mass $s$ such that
\begin{equation}
M=g+s,
\end{equation}
where $M$ is total mass inside the box. The time derivative of
eq.~8 is
\begin{equation}
\dot{M}=\dot{g}+\dot{s},
\end{equation}
and if $\dot{M}$ is taken to be the rate of infall $f(t)$, $\psi(t)$ the star formation rate, and $e(t)$ the mass ejected by stars, then:
\begin{equation}
\dot{s}=\psi(t)-e(t)
\end{equation}
and
\begin{equation}
\dot{g}=f(t)-\psi (t)+e(t).
\end{equation}
The interstellar mass of element $x$ in the zone is $gz_x$, whose time
derivative is:
\begin{equation}
\dot{g} z_x + g \dot{z}_x = -z_x(t) \psi(t) + z^{f}_x f(t) + e_x(t),
\end{equation}
where $z_x(t)$ and $z^{f}_x$ are the mass fractions of $x$ in the
gas and in the infalling material, respectively, and $e_x(t)$ is
the stellar ejection rate of $x$. Solving for $\dot{z}_x(t)$ yields:
\begin{equation}
\dot{z}_x(t)=\{f(t)[z^{f}_x - z_x(t)] + e_x(t) -e(t)z_x(t)\}g^{-1}.
\end{equation}
The second term on the right hand side of eq.~13 accounts for the injection of metals into the gas by stars, while the first and third terms account for effects of dilution due to infall and ejected stellar gas.

In our models we take the infall rate to be:
\begin{equation}
f(t)=\Sigma_{t_o}\left\{\tau_{scale}\left[1-exp\left(-\frac{t_o}{\tau_{scale}}\right)\right]\right\}^{-1}exp\left(-\frac{t}{\tau_{scale}}\right) \ M_{\sun}Gyr^{-1}pc^{-2},
\end{equation}
where $t_o$ and $\Sigma_{t_o}$ are the current epoch and surface density, the latter is taken to be 75M$_{\sun}$pc$^{-2}$, $\tau_{scale}=4$~Gyr, and $t_o$=15~Gyr. This formulation of the infall rate is the one discussed and employed by Timmes, Woosley, \& Weaver (1995) for their models of the Galactic disk.
The rates of mass ejection $e(t)$ and ejection $e_x(t)$ of element $x$ are:
\begin{equation}
e(t)=\int_{m_{\tau_{m}}}^{m_{up}}\{[m-w(m)]\}\psi(t-\tau_m)\phi(m)dm \ M_{\sun}Gyr^{-1}pc^{-2}
\end{equation}
and
\begin{equation}
e_x(t)=\int_{m_{\tau_{m}}}^{m_{up}}\{[m-w(m)]z_{x}(t-\tau_{m})+mp_{x,z_{t-\tau_{m}}}\}\psi(t-\tau_m)\phi(m)dm \ M_{\sun}Gyr^{-1}pc^{-2}.
\end{equation}
In eqs.~15 and 16 $m_{\tau_m}$ is the turn-off mass, i.e. the stellar mass whose main sequence lifetime corresponds to the age of the system; this quantity was determined using results from Schaller et al. (1992). $m_{up}$ is the upper stellar mass limit, taken to be 120~M$_{\sun}$, $w(m)$ is the remnant mass corresponding to ZAMS mass $m$, taken from Yoshii, Tsujimoto, \& Nomoto (1996). $p_{x}(z)$ is the stellar yield, i.e. the mass fraction of a star of mass $m$ which is converted into element $x$ and ejected, and $\phi(m)$ is the initial mass function. The initial mass function is the normalized Salpeter (1955) relation:
\begin{equation}
\phi(m)=\left[\frac{1-b}{m_{up}^{-(1-b)}-m_{down}^{-(1-b)}}\right] m^{-(1+b)},
\end{equation}
where b=1.35.
Finally, the star formation rate $\psi(t)$ is given by:
\begin{equation}
\psi(t)=\nu M \left(\frac{g}{M}\right)^2 M_{\sun}Gyr^{-1}pc^{-2}
\end{equation}
where $\nu=\nu_o\left(1+\frac{z}{0.001}\right)$ is the star formation efficiency, which we note is metallicity-sensitive. Again we have followed Timmes et al. here, but we have introduced the metal-enhancement factor to help fit the data at high O/H. We also note that empirical studies of the star formation rate by Kennicutt (1998) suggest that the SFR exponent is $\sim 1.4 \pm 0.15$ rather than the purely quadratic form in eq.~18. However, test models run with different star formation laws suggested that outcome is much more sensitive to $\nu$ than the exponent, and so we continue to adopt the law given in eq.~18.  

In choosing stellar yields, we assumed at the beginning that carbon and nitrogen levels may be affected by both IMS and massive stars, while oxygen abundance is controlled by massive stars only. Following our analysis in {\S}4,
stellar yields for IMS were taken from VG, while those for massive stars were taken from Maeder (1992). During the numerical calculations, yields were interpolated to find the value relevant for the stellar mass and metallicity under consideration. On the other hand, yields were not extrapolated outside of the ranges for which they were determined by VG and Maeder.
We began by assuming that the VG yields correctly quantify the contributions of IMS to the evolution of these two elements. (We did not experiment with the MBC yields because they have not calculated yields for stars between 5 and 8~M$_{\sun}$.) Then for the contribution of massive stars, we adopted Maeder's (1992) yields for carbon, nitrogen, and oxygen but scaled his carbon yields slightly to force agreement between observations and our model predictions. Our model carbon yields are discussed more in {\S}5.2.1.  

Our calculations assumed a timestep of length one million years for the first billion years, after which the timestep length was increased one hundred-fold. At each timestep, the increment in $z_x$ was calculated by solving eq.~13 along with the required subordinate equations 14-18. This increment was then added to the current value and the program advanced to the next time step. Finally, the total metallicity at each point was taken as the sum of the mass fractions of carbon, nitrogen, and oxygen. 

\subsection{Results Of Numerical Experiments}

Our best numerical model is shown with a bold line and identified with a `B' in Figs.~3A,B along with the data. Models `A' and `C' differ from `B' and each other only in their star formation efficiency and will be discussed later. For model~`B', $\nu_o=0.03$. In both figures this model reproduces the general trends quite well. Note especially the match between the observed NO envelope and model `B' in Fig.~3B. The flat primary, metal-insensitive region below $12+log(O/H)\approx 8.3$ as well as the upward-turning metal-sensitive secondary region at oxygen abundances above this value are reproduced successfully. Given the uncertainties in the observations, model `B' would seem to provide an adequate fit to both the C/O and N/O patterns in Figs.~3A,B. We now discuss these patterns individually.

\subsubsection{Carbon}

The C/O ratio predicted by our model~`B' rises sharply in Fig.~3A as a result of the increase of carbon production and decrease in oxygen production in massive stars, as predicted by Maeder's (1992) yields. This metal-sensitivity of carbon production was strongly implied by our analytical models as well. We note that it was necessary to adjust Maeder's {\it carbon} yields to values somewhat above the published ones in order to achieve a good fit to the data; our final empirical carbon yields for massive stars are listed in Table~3. We point out that these yields have been scaled up by a factor whose value is slightly and positively sensitive to metallicity. (Note that IMS yields would have to have been scaled by more than an order of magnitude to produce the same effect, a change which is much less tenable, given extant planetary nebula abundance constraints.) Column~1 gives the progenitor mass in solar units, columns~2 and 3 compare our and Maeder's yields at $z=0.001$ and columns~4 and 5 do likewise for $z=0.02$. The values listed are in solar masses, i.e. $mp_C(m)$, where $p_C(m)$ is the stellar yield discussed in {\S}3 (see eq.~1). 

In contrast to massive stars, intermediate-mass stars, were found to play an insignificant role in carbon production, based upon the use of VG yields. To confirm this, we recalculated model~B but excluded the IMS carbon contributions. This trial model predicted a track in Fig.~3A only slightly offset from the one shown for model~`B', and so we do not show it. It would seem that IMS have little impact on the model results for carbon. This is interesting because, despite the fact that carbon enrichment in some planetary nebulae definitely indicates carbon production in the progenitors of these objects, it is clear from our analysis, based upon our chosen yields sets, that carbon production by IMS is virtually swamped by an order of magnitude by massive star production of this element. This result was anticipated by our comparison of integrated yields from IMS and massive stars in {\S}3 and Table~2.

\subsubsection{Nitrogen}

By far the most difficult
obstacle to obtaining a reasonable fit to the observations was in reproducing the trends in the nitrogen data. Integrated yields in Table~2 strongly imply that intermediate-mass stars play a major role in nitrogen evolution. Further support for the idea that IMS are an important source of nitrogen is that integrated yields in Table~2 for nitrogen by VG and oxygen by Maeder for $z=0.001$ indicate that $\log (N/O) = -1.41$, in excellent agreement with observed values at low O/H.

But the problem with assuming straightaway that the IMS nitrogen source is the dominant one concerns the delay which these stars undergo in ejecting their products because of their lower masses. Since oxygen is produced by massive stars, a delay in nitrogen ejection by IMS might be too great to explain the N/O value that is observed at low O/H (at ostensibly early times).
Indeed the alleged dominant role of IMS in nitrogen production has been questioned especially by Izotov \& Thuan (1999; their data are indicated with `i' in Fig.~1B), who suggest that because of IMS delay, faster-evolving massive stars must be a significant source of primary nitrogen in order to raise the log(N/O) ratio to a value of -1.5 at 12+log(O/H)$\approx$7.2, a metallicity which they assume to be commensurate with a galactic age too young to allow for nitrogen ejection by IMS. Umeda, Nomoto, \& Nakamura (2000) have suggested that synthesis in metal-free population~III stars can give rise to some primary nitrogen, but their implied production ratios give a log(N/O) that is well below -1.5. 

To analyze this problem, let us first look at the expected delay times for IMS nitrogen ejection by imagining, for simplicity, a star cluster which forms during a burst with mass distributions conforming to a Salpeter IMF.
Figure~4 then follows the increase in the integrated yield (eq.~1) of nitrogen versus time, as predicted by the VG calculations for IMS for the five metallicities indicated by line type. Also shown above the curves with short vertical lines are the main sequence turnoff ages of stars of mass 2, 3, 4, 5, 6, 7, and 8, respectively right to left. The correspondence between age and mass was determined using the results of Schaller et al. (1992).
As the cluster ages, stars of increasingly lower mass eject their products and contribute to the total integrated yield. 
We see in Fig.~4 that nearly all of the nitrogen is produced within roughly 250~Myr of when the population forms, corresponding to the lifetime of a star of 4M$_{\sun}$. Nitrogen yields decline sharply (and the curve flattens) below this mass, because hot bottom burning is less significant. Now to consider the possible significance of this delay time, we compare it with the delay time in the ejection of oxygen by massive stars. Assuming a representative stellar mass of 25M$_{\sun}$, we find that Schaller et al. predict a main sequence lifetime of 6~Myr, considerably shorter than the IMS delay. On short time scales, then, the delay in nitrogen production by IMS appears to be significant, and therefore the suggestion by Izotov \& Thuan that primary nitrogen must come principally from massive stars and its interstellar abundance must increase in lockstep with that of oxygen seems credible.

But the above picture rests on the assumption that the age-metallicity relation for all galaxies is similar. Thurston (1998) has questioned this and has shown that under conditions of a reduced star formation rate, the relation flattens, so that the time required by a system to reach even the lowest metallicity levels observed in H~II regions in metal-poor systems is equal to or greater than the lag time for IMS to begin ejecting nitrogen. Then, assuming continuous star formation thereafter, nitrogen and oxygen will continue to increase in lockstep. 

Bursting systems present a slightly more complicated picture. Yet even here, as long as the time between bursts is significantly larger than the 250~Myr lag time for IMS nitrogen ejection, probability favors observing systems which have advanced at least one lag time beyond the burst, so that by plotting these objects in Fig.~1B we should still find a constant N/O value over a range in metallicity.

Our approach, therefore, was to calculate models with reduced star formation efficiency, i.e. $\nu_o$ in eq.~18, which produced both the N/O and C/O behavior observed in Figs.~1A,B while adopting the yields of VG and Maeder, as discussed in {\S}4, and assuming a significant role of IMS in nitrogen production.
For simplicity, we chose to consider only the continuous star formation scenario and ignore bursting, since we are testing the simplest explanation possible. (An analytical model for time delay of the nitrogen is given in Appendix~B.)

Our best numerical model results for N/O, indicated by curve B in Fig.~3B, provides a good fit to the NO envelope. Notice in particular that the steep rise in log(N/O) at low metallicity coincides with the release of nitrogen by IMS. However, by lowering the star formation efficiency, we have forced this rise to occur at very low oxygen abundances, so that the relatively flat behavior of N/O begins at metallicities below those of the most metal-poor objects observed. At high metallicities, N/O once again rises steeply, this time because nitrogen synthesis in IMS is increasing while oxygen production in massive stars is decreasing with metallicity. Thus, our numerical model shown by curve~B explains the general trend in the NO envelope while preserving the role of IMS in nitrogen production that is expected from the comparison of integrated yields in {\S}3. 

The effect of adjusting the star formation efficiency is shown by curves A and C in Fig.~1B, where efficiency has been lowered and raised, respectively, by a factor of five, as explained at the beginning of {\S}5.2.  As anticipated, the curve enters the plot at a lower metallicity when the efficiency is lower (curve~A); the opposite is seen with the higher efficiency (curve~C). The general shapes of curves A, B, and C are nearly the same at low metallicity, but they are simply shifted horizontally. To explicitly show the time relation here, we have placed five plus symbols corresponding to 0.25, 0.50, 0.75, 1.0, and 2.0~Gyr beginning at the lower left and moving up and to the right on each curve. So, for example, at 0.25~Gyr the value of log~N/O is nearly the same for all three curves, while the metallicity [12+log(O/H)] is clearly different.

Note also that when star formation efficiency is high, as in curve~C, the N/O curve rises more steeply than in the lower efficiency cases of curves~A and B. This result is mainly due to the reduced rate of oxygen buildup at higher metallicities, where the reduced rate occurs for two reasons. First, the massive star oxygen yield decreases with metallicity. Second, the dilution effect of infall is now greater at higher metallicities, since high metallicity is achieved at earlier times, when the infall rate is greater. 
It is interesting that under certain conditions these two effects can conspire to produce a {\it reduction} in the interstellar oxygen fraction, since reduced massive star oxygen production at high metallicities cannot entirely compensate for the effect of dilution by infall. Notice that curve~C does indeed indicate that oxygen abundance briefly goes down (the curve slants slightly leftward) when its 12+log(O/H) reaches a value of about 8.7; the positive trend in oxygen is then resumed when the effects of yield and infall are reversed with metallicity change and time. Such variation may contribute to the scatter observed in the behavior of N/O (see {\S}6), and if stretched far enough might explain the paradoxical situation presently seen in the comparison of the solar metallicity with the significantly lower value in the younger Orion Nebula (Esteban et al. 1998). Of course this regressive behavior of oxygen which is hinted at in our models is heavily dependent on a negative trend of the oxygen yield with metallicity and the existence of a suitable gas flow. The decrease of interstellar metallicity with time would formally require a gas inflow that is a strongly {\it increasing} function of time.

\subsection{Conclusions From Our Numerical Models}
 
Our models strongly indicate that objects located along
the flat region of the N/O curve at low metallicities are entirely and naturally explained if their environment is characterized by historically low star formation rates, with nearly all of the nitrogen being produced by intermediate mass stars. For higher metallicities, the upturn in N/O shows the dependence of nitrogen yield on metallicity, augmented by the decrease in oxygen yield. At the same time, C/O increases with O/H because of the positive metallicity effect on carbon synthesis by massive stars.

\section{Discussion}

We have satisfactorily and simultaneously reproduced the C/O and N/O behavior in Figs.~3A,B by adjusting the star formation rate to suppress oxygen buildup until nitrogen can be ejected by IMS. Our results clearly suggest that carbon rises relative to oxygen because the former is produced in massive stars in a manner which is directly sensitive to metallicity through the mass loss process. On the other hand, the bimodal behavior of N/O is consistent with primary IMS nitrogen production at metallicities below 12+log(O/H) of 8.3, with metal-sensitive secondary production contributing significantly above this value. Both the C/O and high-metallicity N/O trends are amplified by the decrease in oxygen synthesis by massive stars as metallicity increases. This picture is in complete agreement with the yield predictions for massive and intermediate mass stars discussed in {\S}3. 

The low star formation rate at early times is necessary to allow log(N/O) to rise to about -1.4 at metallicities below those observed, as primary nitrogen production for an IMS population builds up to its maximum level. As metallicities continue to climb, IMS production of nitrogen and massive star production of oxygen remain in equilibrium and the element ratio constant. Prior to 250Myr in our model, the gas surface density is roughly equal to 4~M$_{\sun}$/pc$^2$, corresponding to a star formation rate of 10$^{-3}$M$_{\sun}$yr$^{-1}$kpc$^{-2}$.  A reasonable interpretation of this, then, is that the galaxies containing the low metallicity systems studied by Izotov \& Thuan and Kobulnicky \& Skillman (1996) and shown in Fig.~1B historically have had relatively low star formation rates due to their low surface densities. Indeed, results in Papaderos et al. (1996) indicate that a typical surface density for a BCG is 2-5~M$_{\sun}$/pc$^2$, consistent with the surface gas densities in our model at early times. Observations by van~Zee et al. (1997) for low surface brightness galaxies are consistent with this level, as are the total surface densities for the outer disk regions of spirals presented in Vila-Costas \& Edmunds (1992). Of course the star formation rate in these systems could be reduced because of the lack of environmental effects or spiral structure, or the inability by gas to cool efficiently enough. But whatever the cause, it seems clear that the data are consistent with low star formation rates in these objects. 

Thus, our results nicely accomodate the popular picture of blue compact galaxies having an underlying old metal-poor population with star bursts superimposed on the systems. The N/O levels in these systems are set by maximized primary nitrogen production of IMS and oxygen production by massive stars. Our picture contrasts with
the explanation by Izotov \& Thuan which claims that these systems are necessarily very young because of their low metallicities. Then, since their IMS have not had time to begin releasing nitrogen, this forces the conclusion that the N/O ratio at low metallicities must be set by massive star primary nitrogen production. However, this picture neglects the influence of the star formation efficiency in the rate of buildup of metals. According to our analysis,
these systems are simply evolving relatively slowly or intermittently due to historically low star formation rates, so that the delay in nitrogen release by IMS becomes insignificant. 

More generally, any galaxy or galactic region which evolves slowly will maintain a relatively low metallicity over a significant fraction of a Hubble time, since total metallicity is directly related to the star formation rate integrated over time. This opens up the possibility that objects such as blue compact galaxies and outer regions of spiral disks are in fact several Gyr in age despite their low metallicities. This is consistent with the recent conclusions of Legrand (2000), who employed chemical evolution models to study abundances and continuum colors of I~Zw~18. Legrand found that models characterized by a star formation rate of $10^{-4}$ M$_{\sun}$ yr$^{-1}$ over 14 Gyr explain the observations.
This general explanation may also apply to the outer regions of spiral disks, where CNO abundance ratios are similar to those observed in dwarf galaxies such as blue compact objects (Garnett et al. 1999).

Our numerical models allow us to track the separate contributions of IMS and massive stars to carbon and nitrogen evolution. The left and right sides of Fig.~5 show our predictions for carbon and nitrogen, respectively, for our best numerical model, model~B. For each element the fraction of the total mass of carbon or nitrogen being ejected at a point in time is shown in the upper panels for IMS and massive stars versus time. Meanwhile, the lower panels plot the fraction of accumulated mass of carbon and nitrogen ejected by IMS and massive stars, also versus time. 

To gauge the final contributions of IMS and massive stars to carbon and nitrogen synthesis, we can simply read the late-time values off of lower panels in Fig.~5. Clearly, in our models carbon production is dominated by massive stars, with massive stars providing 97\% (all) of the total carbon. On the other hand, the roles of the stellar types appear to be reversed in the case of nitrogen, with IMS providing roughly 90\% (all) of this element.

{\it Our analysis has led to the conclusion that carbon and nitrogen production in the Universe are essentially decoupled from one another, with the former produced in massive stars, while the latter is produced in IMS.} Consensus toward this claim in the case of carbon has been building for several years. For example, work by Prantzos et al. (1994), Garnett et al. (1999), and Gustafsson et al. (1999) show that carbon is largely produced by massive stars, to which we now add our support. In the case of nitrogen, the predicted large IMS yields in conjunction with the large extant database of planetary nebula abundances suggest strongly that the production of this element is dominated by IMS. The most recent problem has been the one presented by the observed constant N/O value at low metallicities, suggesting that massive stars must play at least an important role here, due to supposed significant IMS nitrogen release delays. However, our models have shown that low star formation rates can accomodate IMS nitrogen production with no problem.

Finally, during this investigation we have chosen to ignore until now the large scatter in N/O that is observed at a single O/H value. This matter has been addressed by a number of authors including Garnett (1990), Pilyugin (1993; 1999), and Marconi, Matteucci, \& Tosi (1994), with the consensus being that at least some of the scatter is real and due to bursts which momentarily lower N/O in the observed H~II regions with sudden injections of fresh oxygen, but as IMS eject nitrogen after the customary lag time, N/O rises again. Therefore, the scatter comes about by observing a large sample of H~II regions in various stages of oxygen and nitrogen enrichment.
This picture implies that most points should be concentrated at relatively high N/O values, with fewer points, representing those objects experiencing sudden oxygen enrichment, located below the main concentration, since presumably bursts are followed by relatively long periods of quiescence during which the end point N/O is characteristic.

However, a close look at the vertical distribution of points in N/O in Fig.~1B reveals that most points seem to be clustered along the NO envelope at relatively low values with the concentration falling off as one considers higher N/O values, i.e. exactly the reverse of the standard interpretation of the scatter described above. Indeed, this empirical finding strongly suggests that the ``equilibrium'' or unperturbed locus where most H~II regions reside is the NO envelope, and thus the excursions caused by sudden injections of material are actually {\it upward} toward the region of fewer points.

Barring an unidentified selection effect, then, the distribution of data points in Fig.~1B would seem to challenge the conventional picture. In fact the data are consistent with the lack of evidence for localized oxygen contamination from massive stars in H~II regions described in Kobulnicky \& Skillman (1997b). 
Furthermore, the falloff in points above the NO envelope is more consistent with injections of nitrogen rather than oxygen; in this case the nitrogen source might be Wolf-Rayet stars or luminous blue variable stars, both of which were considered by Kobulnicky \& Skillman (1997a) in their study of nitrogen-enriched H~II regions in NGC~5253. Their explanation suggests a simultaneous enrichment of helium, and thus H~II regions exhibiting high values of N/O in Fig.~1B should be checked for evidence of helium enrichment. Our general results, though, imply that the contributions of WR and luminous blue variable stars to nitrogen enrichment must be small.
We postpone further investigation of this aspect of the N/O distribution, as we plan to take it up in another paper.

Finally, we note the upper limits of log(N/O) for four damped Ly$\alpha$ systems taken from Lu et al. (1996). While two of the points lie within the NO envelope, the other two points lie about 0.25 dex below it. Since the position of the rising track of a numerical model in Fig.~3B can be forced to the right (higher star formation rates) and left (lower star formation rates), such objects and others at even lower N/O values (if eventually observed) may be explained as representing very early stages of N/O evolution when IMS are just beginning to release nitrogen, i.e. roughly 250 Myr or less after an initial star burst. 

\section{Summary}

We have compiled and analyzed several sets of published stellar yields for both massive and intermediate mass stars. Using analytical models with accretion but no time delay we have chosen yields which seem most plausible for explaining the data and have employed these yields in numerical models which predict the buildup of carbon, nitrogen, and oxygen over time. Our analysis suggests the following:

\begin{enumerate}

\item The most appropriate yields appear to be those of van~den~Hoek \& Groenewegen (1997) for intermediate-mass stars and Maeder (1992) for massive stars.

\item Carbon is produced mainly by massive stars, with at most only a slight contribution from intermediate-mass stars. Observations of C/O versus O/H are consistent with a metallicity-enhanced carbon yield from massive stars, in agreement with Maeder's (1992) predictions.

\item Nitrogen, conversely, is produced principally in intermediate-mass stars, specifically those stars between 4 and 8~M$_{\sun}$ which undergo hot bottom burning and expel large amounts of primary nitrogen at low metallicities, and secondary/tertiary nitrogen at higher z levels. This conclusion agrees with the relatively large size of the nitrogen yields of intermediate mass stars predicted by van~den~Hoek \& Groenewegen (1997).

\item Carbon and nitrogen production appear to be essentially decoupled from one another, since they are produced in two entirely different sites, i.e. massive stars and intermediate-mass stars.

\item The characteristic delay in nitrogen release by intermediate mass stars is approximately 250~Myr.

\item Time delay for both carbon and nitrogen production does not appear to be an important factor in the evolution of these elements.

\item Oxygen is produced entirely by massive stars. Maeder's (1992) oxygen yields, which decrease with increasing metallicity, are consistent with observed behavior of C/O and N/O.

\item The observed behavior of N/O versus O/H is bimodal and is related to the primary/secondary nature of nitrogen production as well as the inverse sensitivity of oxygen yields to metallicity. At low metallicities the relation is flat since, in this region, nitrogen is primary in origin and simply increases in lockstep with oxygen. Beginning at roughly 12+log(O/H)=8.3, secondary/tertiary nitrogen production becomes significant and produces the upturn in the data seen at higher metallicities. This same upturn is augmented by the decrease in massive star oxygen production at higher metallicities.

\item Nitrogen production at low metallicities is consistent with production in intermediate mass stars despite the delay which these stars experience in the release of their nitrogen. Since the relevant stars have progenitor masses between 4-8~M$_{\sun}$ and maximum lifetimes of roughly 250~Myr, a low star formation rate and the consequent slow rise in metallicity, is consistent with the data.

\item Our findings allow for objects such as blue compact galaxies and regions such as the outer parts of spiral disks, all characterized by low metallicity, to nevertheless have formed many Gyr ago.

\end{enumerate}

Finally, because of the relatively short delay in nitrogen release by intermediate-mass stars, it may not be possible to employ N/O as a useful galaxy age indicator (as proposed by Edmunds \& Pagel 1978, a time delay of 1 to 2 Gyr or so would be needed), except perhaps for extremely young systems. Nevertheless, we believe that our identification of the important C and N sources now allows a reasonable explanation of the systematics of the behavior of C/O and N/O ratios in galactic systems.

Future work should include a closer look at the N/O scatter by performing a consistent abundance determination analysis on the various data in the literature. Analysis should include consideration of H~II region structure and excitation, and other means for explaining scatter. The possibility that the scatter is related to nitrogen contamination from Wolf-Rayet or luminous blue variable stars should be investigated further, although our general results imply that their impact on the overall yield of nitrogen from a stellar population must be small.

\acknowledgments

R.B.C.H would like to thank the members of the Physics \& Astronomy Department of Cardiff University, Wales, for their hospitality and support during an extended visit. Travel to Cardiff was made possible by a generous travel grant from the University of Oklahoma. We also thank Bernard Pagel, Trinh Thuan, and Evan Skillman for useful and informative discussions.

\appendix

\section*{Appendix A}

\section*{Limiting Analytical Models}

We give here additional limiting forms of the analytic equations from {\S}4. 

Carbon and nitrogen abundances in the ``simple'' closed-box model, with no time 
delays (limiting forms of eqs.~5 and 7):
\begin{equation}
\eqnum{A1}
z_c=\frac{p_{pc}}{p}z+\frac{p_{sc}}{2p}z^2
\end{equation}
\begin{equation}
\eqnum{A2}
z_n=\frac{p_{pn}}{p}z+\frac{p_{sn}p_{pc}}{2p^2}z^2+\frac{p_{sn}p_{sc}}{6p^2}z^3
\end{equation}

\section*{Appendix B}

\section*{Analytical Models With Time Delay}

To give some indication of the effect of time delays on the element ratios we 
now give an elementary model. This must inevitably make assumptions about the 
time evolution of star formation, so the model is not general. For convenience 
(and to avoid an epidemic of parameters) we neglect infall and just consider the 
simple closed-box model (although the model can be straightforwardly extended to 
include accretion). Suppose that all primary nitrogen is 
delayed by a time $\tau_n$ and then suddenly released. Let the star 
formation rate in the closed system be simply 
proportional to gas density (with a constant $k$), total mass of the system be 
unity, starting from just unenriched gas at $t=0$, then:
\begin{equation}
\eqnum{B1}
\frac{ds}{dt}=k \alpha g=k \alpha e^{-\alpha kt},
\end{equation}
implying $g=e^{-\alpha kt}$, $s=1-e^{-\alpha kt}$, and $z=p\alpha kt$, so that 
overall metallicity (including oxygen) increases linearly with time. This allows 
us to map time into metallicity, and {\it vice versa}. We may represent the time 
delay as a metallicity increment $z_{\tau_n}=p\alpha k\tau_{n}$. Now $(\alpha 
k)^{-1}$ represents the timescale for star formation  (see eq.~B1), so if we 
denote $\tau_{\ast}=(\alpha k)^{-1}$, then 
$\frac{z_{\tau_n}}{p}=\frac{\tau_n}{\tau_{\ast}}$, and choosing different values 
of $\frac{z_{\tau_n}}{p}$ therefore corresponds to different rates of star 
formation for the system. We define a step function
\begin{eqnarray*}
\eqnum{B2}
H(x-x_o)&=&1\ \mbox{if}\ x\ge 0\\
&=&0\ \mbox{if}\ x<0
\end{eqnarray*}
and then 
\begin{eqnarray}
\eqnum{B3}
z_n&=&\frac{p_{pn}}{p}\exp\left(\frac{z_{\tau_n}}{p}\right)(z-z_{\tau_n})H(z-z_{
\tau_n})\\
&&+\frac{p_{sn}p_{pc}}{2p^2}z^2+\frac{p_{sn}p_{sc}}{6p^2}z^3\nonumber
\end{eqnarray}
We can see the effect of a time delay on the primary nitrogen in a system 
with fast star formation by setting $z_{\tau_n}=0.01$. This is plotted as 
a dashed line in Fig.~A1B.

\section*{Appendix C}

\section*{Analytic Model of Variable Oxygen Yield}

We show here that our general conclusions about the sources of carbon
and nitrogen are not affected by the possibility that the oxygen yield
is a function of metallicity. The effect of a decrease with increasing
metallicity is to steepen
the relations of C/O and N/O with O/H at high O/H, but not to introduce 
any qualitatively new effects.

Suppose that the yield for oxygen has the form $p_{1}$-$p_{2}$$z_{o}$, noting
the minus sign, $p_{2}$ is positive, then for a simple closed-box model
with no accretion we have:

\begin{equation}
\eqnum{C1}
z_c = -\frac{1}{p_2}\left(p_{pc} + \frac{p_{sc}p_1}{p_2}\right)\ln \left(1-\frac{p_2 z_o}{p_1}\right) - \frac{p_{sc}z_o}{p_2}
\end{equation}
and
\begin{equation}
\eqnum{C2}
z_n = -\frac{1}{p_2}\left(p_{pn} - \frac{p_{sc}p_{sn}p_1}{p_2^2}\right) ln\left(1-\frac{p_2z_o}{p_1}\right) + \frac{p_{sn}}{2p_2^2}\left(p_{pc} + \frac{p_{sc}p_1}{p_2}\right) ln^2\left(1-\frac{p_2z_o}{p_1}\right) + \frac{p_{sc}p_{sn}z_o}{p_2^2}
\end{equation}

These are plotted in Fig.~A1A,B, with $p_{1}$=0.017, $p_{2}$=0.5, similar to the 
Maeder models. The steepening effect can be seen, but there is no 
incentive to revise our basic conclusions on the sources of C and N.

\clearpage

\begin{deluxetable}{lccc}
\tablecolumns{4}
\tablewidth{0pc}
\tablenum{1}
\tablecaption{Summary of Observations}
\tablehead{
\colhead{Object}&\colhead{Author\tablenotemark{1}}&\colhead{C/O}&\colhead{N/O}
}
\startdata
H II (OPT,RAD)&Shaver&\nodata&20\\
H II (FIR)&Afflerbach&\nodata&34\\
H II (OPT)&V{\'i}lchez&\nodata&9\\
H II (OPT)&Fich&\nodata&4\\
H II (FIR)&Rudolph&\nodata&5\\
H II (OPT) & Kobulnicky &3&70\\
H II (OPT) & van~Zee &\nodata&173\\
H II (OPT) & Thurston&\nodata&165\\
H II (OPT) & Izotov &10&53\\
H II (OPT) & Garnett & 14 & \nodata\\
DLA & Lu et al. &\nodata&5 \\
FG Stars & Gustafsson & 54 & \nodata \\
B Stars & Gummersbach & 16 & 16 \\
Halo Stars & Tomkin & 34 & \nodata 
\enddata
\tablenotetext{1}{Shaver et al. (1983);
Afflerbach et al. (1997); V{\'i}lchez \& Esteban (1996); Fich 
\& Silkey (1991); Rudolph et al. (1997); Kobulnicky: Kobulnicky \& 
Skillman (1996, Nitrogen; 1998, Carbon); van~Zee et al. (1998); 
Thurston, Edmunds, \& Henry (1996); Izotov \& Thuan (1999); 
Garnett et al. (1995, 1997, 1999); Lu et al. (1996); Gustafsson et al. (1999); Gummersbach et al. (1998); Tomkin et al. (1992)}  
\end{deluxetable}

\clearpage

\begin{deluxetable}{lcccccc}
\tablecolumns{7}
\tablewidth{0pc}
\tablenum{2}
\tablecaption{Integrated Yields, P$_x$\tablenotemark{1}}
\tablehead{
\colhead{Source\tablenotemark{2}} &
\colhead{m$_{d}$} &
\colhead{m$_{u}$} &
\colhead{z\tablenotemark{3}} &
\colhead{P$_C$} &
\colhead{P$_N$} &
\colhead{P$_O$} 
}
\startdata
VG &1&8&.001&6.7E-4&5.5E-4&\nodata \\
VG &1&8&.004&7.2E-4&5.8E-4&\nodata \\
VG &1&8&.008&6.1E-4&6.4E-4&\nodata \\
VG &1&8&.02&4.1E-4&7.4E-4&\nodata \\
VG &1&8&.04&1.1E-4&9.4E-4&\nodata \\
MBC&1&5&.008&2.1E-3&1.8E-4&\nodata \\
MBC&1&5&.02&3.4E-4&2.0E-4&\nodata \\
M&9&120&.001&1.6E-3&2.8E-6\tablenotemark{4}&1.6E-2 \\
M&9&120&.02&7.7E-3&2.9E-4\tablenotemark{4}&5.6E-3 \\
WW&11&40&0.0&6.7E-4&1.1E-6&3.5E-3 \\
WW&11&40&2E-6&7.6E-4&5.5E-8&5.7E-3 \\
WW&11&40&2E-4&7.3E-4&2.6E-6&6.0E-3 \\
WW&11&40&.002&7.4E-4&2.4E-5&6.3E-3 \\
WW&11&40&.02&8.1E-4&2.6E-4&7.0E-3 \\
N&13&70&.02&4.9E-4&9.1E-6&1.3E-2 
\enddata
\tablenotetext{1}{Mass fraction of all stars formed
which is eventually expelled as new $x$ from stars in the range m$_{d}$ and $m_{u}$}
\tablenotetext{2}{VG=van~den~Hoek \& Groenewegen (1997); MBC=Marigo, 
Bressan, \& Chiosi (1996; 1998); M=Maeder (1992); WW=Woosley \& Weaver (1995); 
N=Nomoto et al. (1997).}
\tablenotetext{3}{Metallicity mass fraction, where z$_{\sun}$=0.02}
\tablenotetext{4}{These values do not include contributions from supernova ejecta, only winds; they represent lower limits}  
\end{deluxetable}

\clearpage

\begin{deluxetable}{lcccc}
\tablecolumns{5}
\tablewidth{0pc}
\tablenum{3}
\tablecaption{Comparison Of Carbon Yields{\tablenotemark{1}}\ \ For Massive Stars}
\tablehead{
Stellar & \multicolumn{2}{c}{z=0.001} & \multicolumn{2}{c}{z=0.02} \\ \cline{2-3}\cline{4-5}
\colhead{Mass(M$_{\sun}$)} & \colhead{Model} & \colhead{Maeder} & \colhead{Model} & \colhead{Maeder}
}
\startdata
9&9.9E-2&5.6E-2&5.2E-2&2.7E-2 \\
12&1.8E-1&1.0E-1&1.3E-1&7.1E-2 \\
15&3.4E-1&2.0E-1&2.6E-1&1.4E-1 \\
20&5.2E-1&3.0E-1&4.1E-1&2.2E-1 \\
25&7.1E-1&4.0E-1&5.6E-1&3.0E-1 \\
40&9.7E-1&5.5E-1&9.0&4.9 \\
60&1.2&7.0E-1&1.3E+1&7.2 \\
85&1.3&7.2E-1&2.5E+1&1.3E+1 \\
120&1.5&8.8E-1&1.5E+1&8.0 
\enddata
\tablenotetext{1}{Stellar yield $mp_C$ in M$_{\sun}$ for metallicity indicated, where $p_C$ is the mass fraction of stellar mass $m$ which is converted to new $C$ and expelled}
\end{deluxetable}

\clearpage

\section*{Figure Captions}

\figurenum{1A}\figcaption{log(C/O) versus 12+log(O/H) for extragalactic H~II
regions and stars.  H~II region data are from Garnett et al.  (1995, 1997, 1999, G), Izotov \& Thuan (1999; I), and Kobulnicky \&
Skillman (1998, K). The symbols M and S show the positions of the Galactic H~II region
M8 and the sun, respectively.  The filled circles show F and G stellar data from
Gustafsson et al. (1999), the filled boxes are B star data from Gummersbach et al. (1998), and the filled diamonds are halo star data from Tomkin et al. (1992).  Typical uncertainties are shown in the upper left.}

\figurenum{1B}\figcaption{log(N/O) versus 12+log(O/H) for H~II
regions and stars in the Milky Way disk, extragalactic spirals, and
irregulars. Data for the Milky Way are from Afflerbach et al.
(1997, A); Fich \& Silkey (1991, F); Shaver et al. (1983, S);
Rudolph et al. (1997, R); and V{\'i}lchez \& Esteban (1996, V).
Extragalactic data are from Izotov \& Thuan (1999, i); Kobulnicky
\& Skillman (1996, K); Thurston et al. (1996, T); and van~Zee et al.
(1998, Z). Filled circles are stellar data from Gummersbach et al. (1998). The circle indicates the position of the Orion Nebula
(Esteban et al. 1998), the large S shows the position of the sun
(Grevesse et al. 1996), and the L symbols at extremely low oxygen
show upper limits for two high redshift damped Lyman-$\alpha$
objects in Lu et al.
(1996).}

\figurenum{2A}\figcaption{log(C/O) vs. 12+log(O/H) for analytical model predictions. Data points are same as those in Fig.~1A. The bold solid line shows our best fit analytical model with $a=0.1$, while the light solid lines show the same models but with $a=0.5$ and $a=0.9$. Values of $a$ less than 0.1 produce curves which are indistinguishable from the $a=0.1$ curves, so we do not show them.} 

\figurenum{2B}\figcaption{Same as Fig.~2A but for log(N/O) vs. 12+log(O/H).}

\figurenum{3A}\figcaption{log(C/O) vs. 12+log(O/H) for numerical model predictions. Data points are same as those in Fig.~1A. Curve~B is our best model with $\nu_o = 0.03$, while curves~A and C show the impact of reducing and increasing, respectively, the star formation efficiency by a factor of five, i.e. $\nu_o = 0.006$ for A and $\nu_o = 0.15$ for C.}

\figurenum{3B}\figcaption{Same as Fig.~3A but for N/O. The plus symbols show time points, starting at lower left, for 0.25, 0.50, 0.75, 1.0 and 2.0~Gyr.}

\figurenum{4}\figcaption{Total nitrogen mass fraction synthesized and ejected by a coeval population of intermediate mass stars as a function of time. For any time on the abscissa, the ordinate value is the mass fraction ejected by all stars up to that time. Vertical lines show main sequence turnoff times for solar masses 2, 3, 4, 5, 6, 7, and 8, right to left, respectively. Progenitor star metallicities are identified by line types, as indicated in the legend.}

\figurenum{5}\figcaption{Left Panels: Fraction of synthesized carbon being ejected instantaneously as a function of time (top) and fraction of total carbon synthesized and ejected up to a time (bottom) as functions of time in Gyr for IMS and massive stars. Right Panels: Same as left panels but for nitrogen. All lines in this figure refer to results from our model~B, described in {\S}5.2.}

\figurenum{A1A}\figcaption{Same as Fig.~1A but showing our best-fit analytical accretion model from {\S}4 plus a simple model with metal-sensitive oxygen yield (see Appendix C) and $p_2$ set at 0.5.}

\figurenum{A1B}\figcaption{Same as Fig.~1B but showing our best-fit analytical accretion model from {\S}4 plus a simple model with metal-sensitive oxygen yield (see Appendix C) and $p_2$ set at 0.5. The dashed line shows the results of a time-delay model described in Appendix~B.}

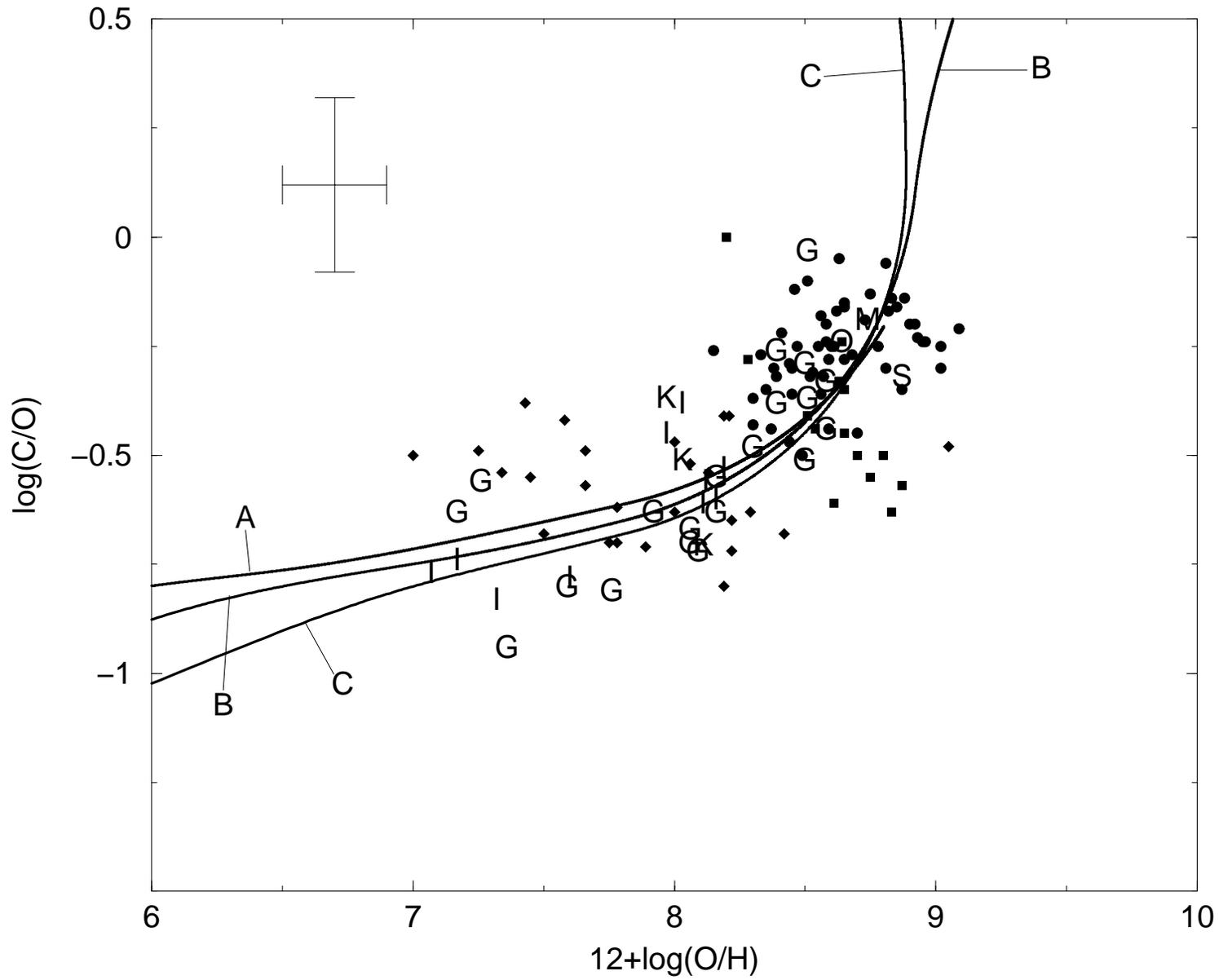
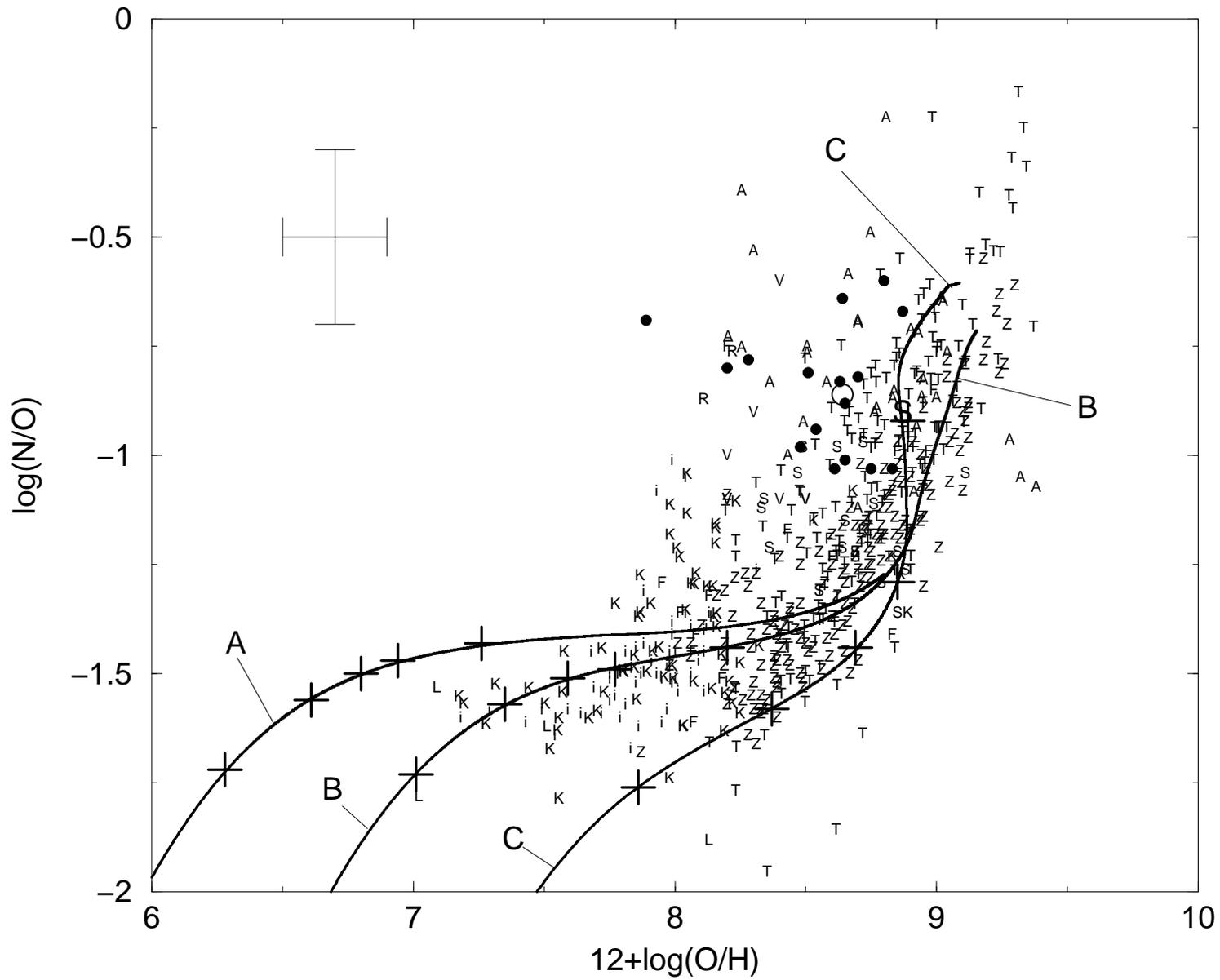
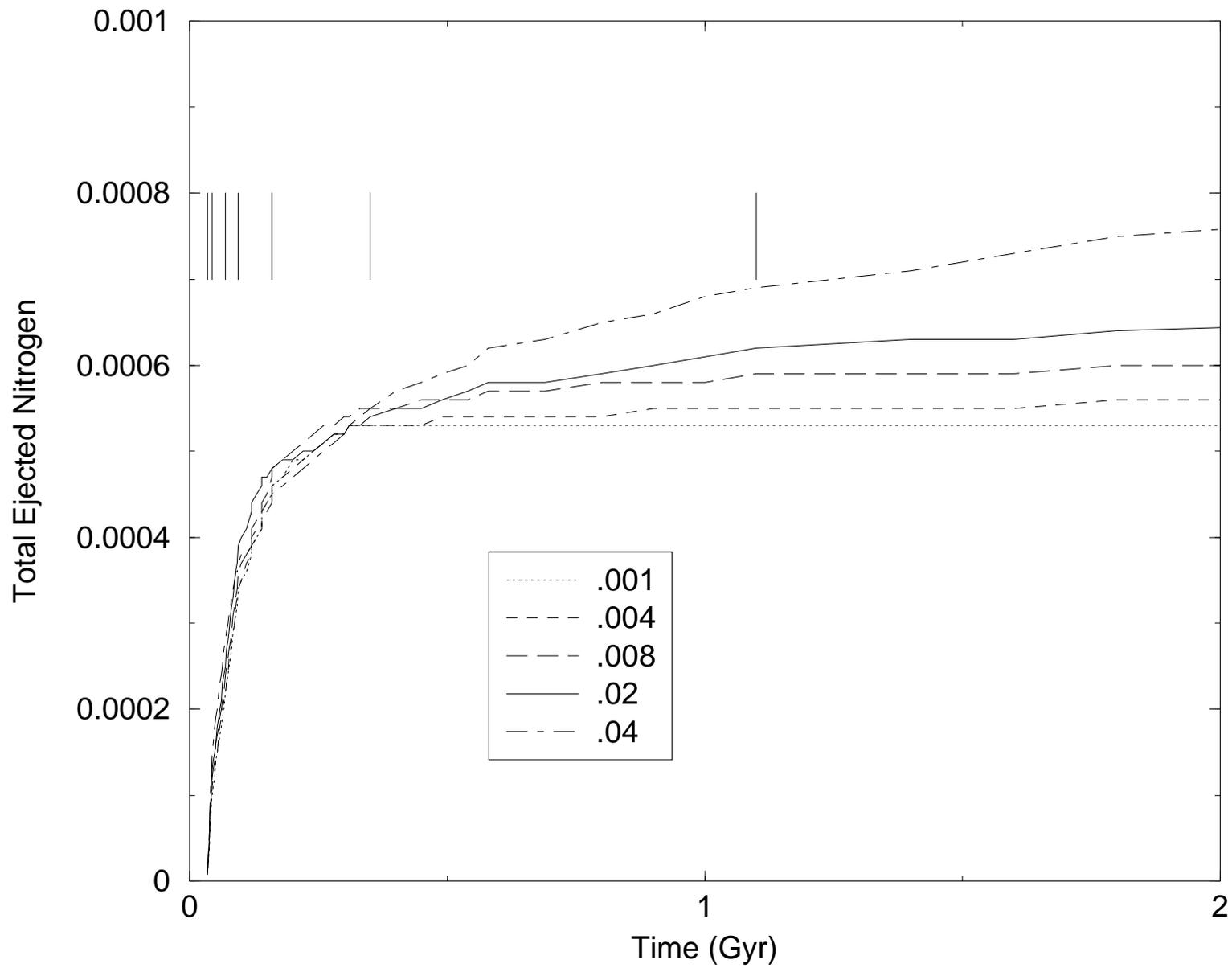
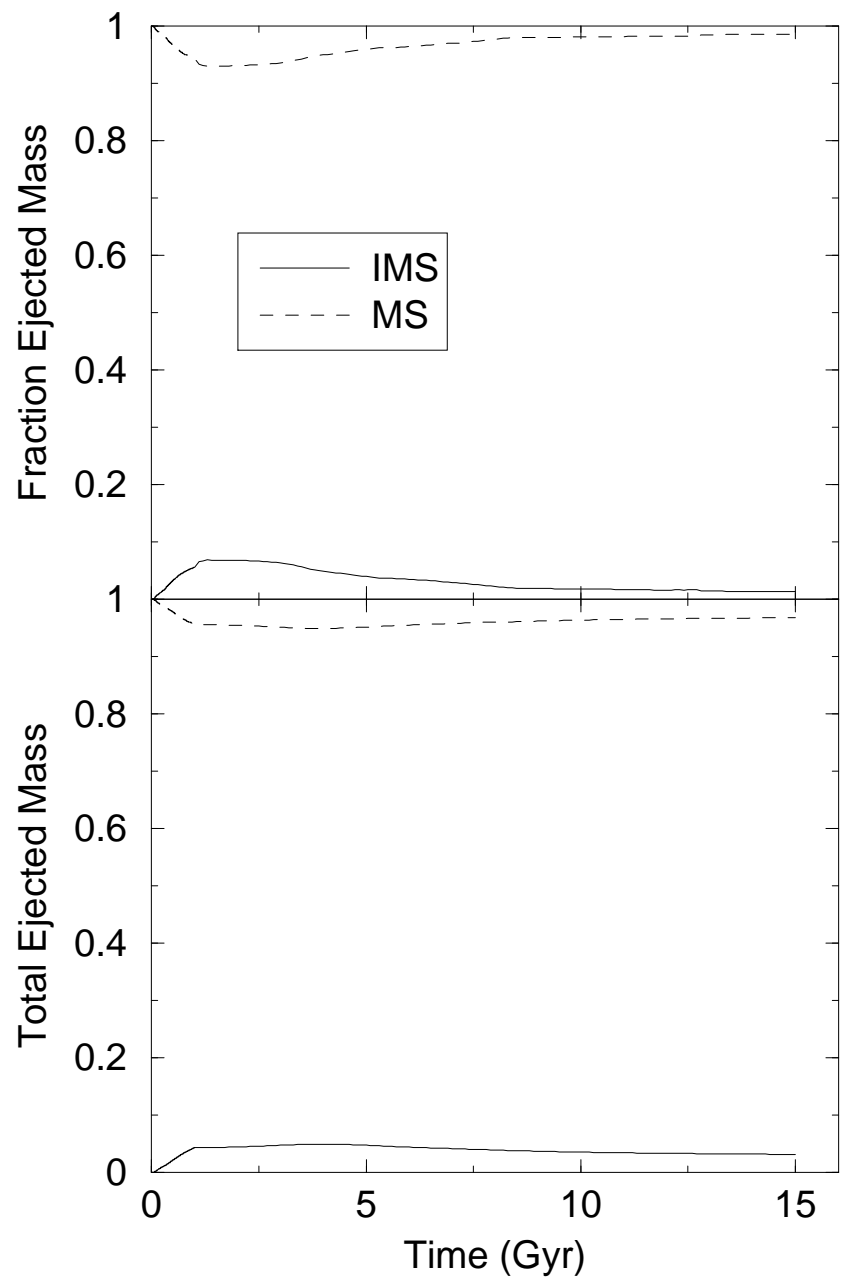
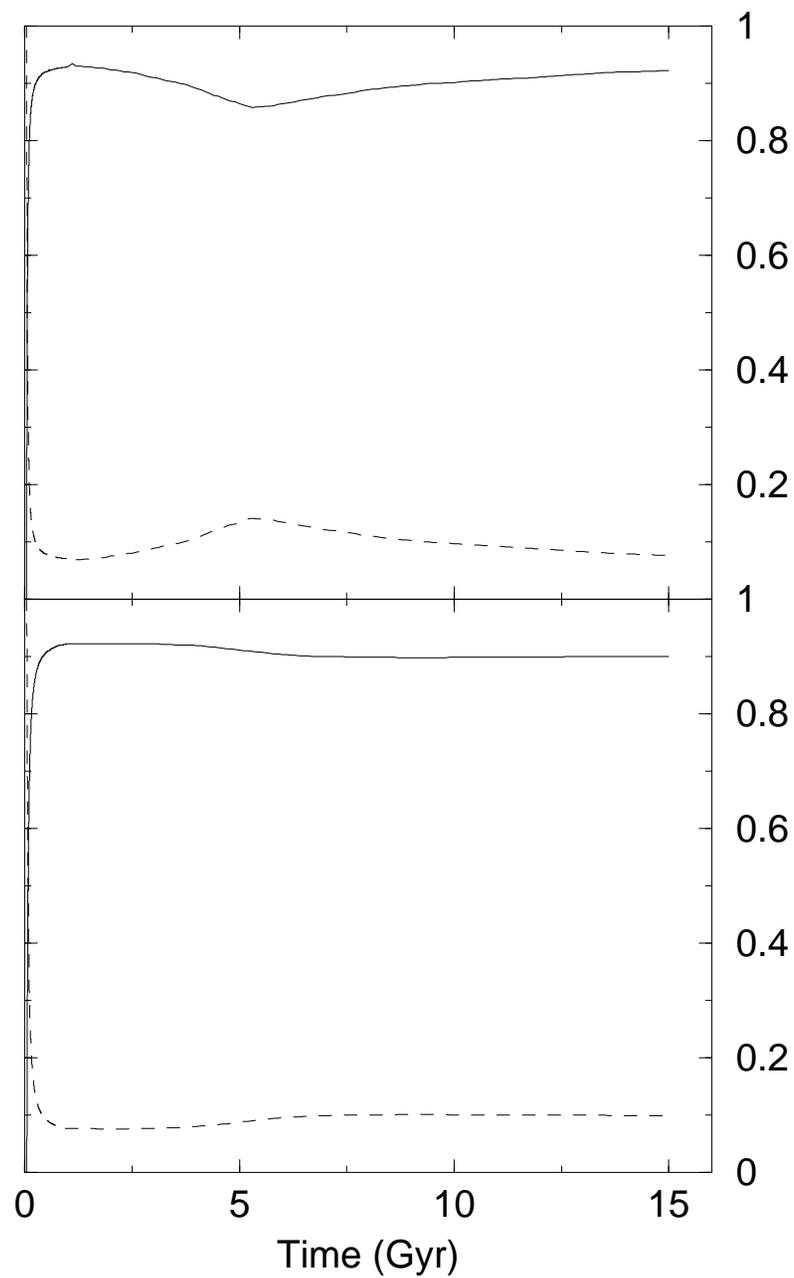
\end{document}